\documentclass[%
 superscriptaddress,
 preprint,
 showpacs,
 amsmath,
 amssymb,
 aps,
 pra,
 showkeys
]{revtex4-1}

\usepackage{graphicx}

\begin{document}

\title{Optimal control theory for quantum electrodynamics: an initial state problem}

\author{Alberto Castro}
\affiliation{ARAID Foundation, Av. de Ranillas 1-D, 50018 Zaragoza (Spain)}
\affiliation{Institute for Biocomputation and Physics of Complex Systems,
University of Zaragoza, Calle Mariano Esquillor, 50018 Zaragoza (Spain)}
\email{acastro@bifi.es}

\author{Heiko Appel}
\affiliation{Max Planck Institute for the Structure and Dynamics of Matter and Center for Free-Electron Laser  Science, Luruper Chaussee 149, 22761 Hamburg, Germany}

\author{Angel Rubio}
\affiliation{Max Planck Institute for the Structure and Dynamics of Matter and Center for Free-Electron Laser  Science, Luruper Chaussee 149, 22761 Hamburg, Germany}
\affiliation{Center for Computational Quantum Physics (CCQ), Flatiron Institute,  New York NY 10010}
\affiliation{Nano-Bio Spectroscopy Group, Universidad del Pa{\'{\i}}s Vasco, 20018 San Sebastián, Spain}

\date{\today}



\begin{abstract}
  In conventional quantum optimal control theory, the parameters
  that determine an external field are optimised to
  maximise some predefined function of the trajectory, or of the
  final state, of a matter system. The situation changes
  in the case of quantum electrodynamics, where the
  degrees of freedom of the radiation field are now part of the
  system. In consequence, instead of optimising an external
  field, the optimal control question turns into a optimisation
  problem for the many-body initial state of the combined
  matter-photon system. In the present work, we develop such a optimal
  control theory for quantum electrodynamics. We derive the equation
  that provides the gradient of the target function, which is often the
  occupation of some given state or subspace, with respect to the
  control variables that define the initial state. We choose the
  well-known Dicke model to study the possibilities of this
  technique. In the weak coupling regime, we find
  that Dicke states are the optimal matter
  states to reach Fock number states of the cavity mode with large
  fidelity, and vice versa, that Fock number states of the photon modes
  are the optimal states to reach the Dicke states. This picture does not
  prevail in the strong coupling regime. We have also considered the extended
  case with more than one mode. In this case, we find that increasing
  the number of two-level systems allows to reach a larger occupation
  of entangled photon targets.
\end{abstract}

\maketitle

\section{\label{sec:introduction} Introduction}


The coherent control of the dynamics of quantum
systems~\cite{Shapiro2003} is frequently theoretically approached
within the framework of quantum optimal control theory~\cite{Brif2010}
(QOCT), the subset of optimal control theory~\cite{Kirk1998}
applicable to quantum processes. It is concerned with the
identification of parameters of the Hamiltonian of a system that
induce a pre-defined optimal behaviour, such as the maximal occupation
of an excited state, the dissociation of a molecule, etc. Often, those
parameters -- the \emph{control variables} -- determine the temporal
shape of an external field present in the Hamiltonian -- the
\emph{control function}. For example, the control problem may be the
identification of some laser field shape (described classically) that
prepares a quantum system in a desired state.

This formulation assumes that the controlling field is external, and
its source is not affected by the system evolution. If, instead, one
needs to describe the coupled evolution of a piece of matter (atom,
molecule, etc.) and an electromagnetic field, the dynamical system
must in fact be composed of both. This is the situation, for example,
in the field of cavity quantum electrodynamics
(CQED)~\cite{Walther2006} or in quantum optics in general.  An atom or
molecule trapped in a cavity and the electromagnetic cavity radiation
form, in principle, a closed system, and no external control can be
exercised: once the initial state is set, the evolution is fixed. Of
course, one may then introduce an additional and external classical
field, or a temporal variation of the system couplings, that can be
controlled in some way. In this way, the formulation of the control
problem would still be the traditional one~(see
Refs.~\cite{Rojan2014,Deffner2014,Allen2017}, for example).


However, one may be interested in finding the optimal cavity field
that induces a given matter subsystem behaviour. In this case, the
control formulation apparently changes, as the control variables
should be set to define the initial state of the field: the task is to
find the initial field state such that the coupled evolution is
optimal with respect to the goal. And, once the field and the matter
subsystem are treated on the same footing, one may pose an
\emph{inverse} control problem: find the optimal matter system state
that induces the creation of a given photon field. In any case, the
task has changed: from manipulating parameters that determine the
Hamiltonian (or, in general, the dynamical function), to manipulating
parameters that determine the initial state.  Fortunately, it is known
that the two problems are mathematically exactly equivalent. Using
this formal equivalence, we will present working equations for the
\emph{optimal-initial-state} problem. Note that the two points of
  view are not exclusive, and one may use the formalism described
  below also for open systems subject to external fields -- or subject
  to dissipation.

The objective of this work is to explore, numerically, the control
technique in this \emph{apparently} new formulation. For that purpose
we have focused on a relevant quantum optical model: the Dicke
Hamiltonian~\cite{Dicke1954,Garraway2011}, that describes a set of
two-level-systems (TLS) coupled to a single cavity mode -- or more
than one mode, in extended versions of the model. Originally, it was introduced to
explain superradiant emission. Although hard to realise
experimentally, there is interest in it because of the possibility of
creating very entangled matter and photon states, and because of its phase
transitions.

For the present work, we have designed the following numerical experiments: (1)
find the optimal initial photon field such that the evolution of the
system, assuming the TLS set enters the cavity in its ground state,
produces a Dicke or $W$ state~\cite{Dicke1954,Dur2000}; and,
inversely, (2) find the optimal TLS set initial state such that the
evolution of the system, assuming a vacuum initial photon field,
produces a photonic Fock state -- or a combination of Fock states in a
multi-mode setup.

Section~\ref{sec:oct} presents the basic theoretical discussion
underlying the following calculations; Section~\ref{sec:model}
introduces the physical model, and the relevant states;
Section~\ref{sec:results} describes the optimisation results, and
finally conclusions are summarised in Section~\ref{sec:conclusions}.

\section{\label{sec:oct} Optimal Control versus Initial State Preparation}

We start by stating a simple formulation of the usual (quantum)
optimal control
problem~\cite{Brif2010,werschnik2007,peirce1988,Tersigni1990,Castro2012}. Let $y$
be a Hilbert state vector that evolves according to the evolution law:
\begin{eqnarray}
\label{eq:ysys1}
\dot{y}(t) & = & - {\rm i}\hat{H}(u,t)y(t) + b(u,t)\,,
\\
\label{eq:ysys2}
y(0) & = & y^0\,.
\end{eqnarray}
The Hamiltonian $\hat{H}(u,t)$ depends on a set of parameters, the
control variables $u_1,\dots,u_m$. Note, that we have added an extra
inhomogeneous term $b(u,t)$ that will become useful below, although it
is obviously not present in a normal Schr{\"{o}}dinger equation. More
general quantum evolutions could be considered, e.g. open systems
through the incorporation of stochastic terms, or via the use of
Lindblad equations. Sometimes, instead of a discrete set of control
variables $u_1,\dots,u_m$ one considers one or more continuous
\emph{control functions} $u(t)$ as the objects of optimisation~\footnote{Note
that there is no loss of generality in considering one or the other
option: in any numerical implementation of the problem, each function
is represented by a set of parameters; inversely, any set of
parameters can be considered as a set of functions that are
constrained to be constant in time.}.

The system evolves until time $T$. One may then define a performance
measure, or target functional $J$:
\begin{equation}
\label{eq:targetfunctional}
J(y,u) = F(y(T),u) + \int_0^T\!\!{\rm d}t\; G(y(t),u,t)\,,
\end{equation}
whose maximisation is defined as the optimal system behaviour. In
general it may contain the two terms shown in
Eq.~(\ref{eq:targetfunctional}). The first term only depends on the
final state of the system, whereas the second one depends on the full
trajectory.  In the following, we will only consider the presence of
the first term $F$.

The specification of a control $u$ determines the system trajectory,
$u \to y_u$. The optimal control problem may then be formulated as the
identification of the maximum of the function 
\begin{equation}
\label{eq:g1}
G(u) = F(y_u(T),u)\,.
\end{equation}

Often, the control variables $u$ parametrize the shape of an external
field; the problem is therefore finding the best external field that
induces a given system reaction. However, if the field cannot be
considered to be external, but must be included as a part of the full
state description, the evolution is fixed once the initial state is
specified. Finding the best external field means in this case shaping
its initial state. The control problem must then be formulated for a
system with the form:
\begin{eqnarray}
\label{eq:zsys1}
\dot{z}(t) & = & -{\rm i}\hat{H}z(t)\,,
\\
\label{eq:zsys2}
z(0) & = & z^0(u)\,,
\end{eqnarray}
or, more compactly, $z(t) = \hat{U}(t,0)z^0(u)$, where
$\hat{U}(t_1,t_2) =$ $\exp(-{\rm i}(t_1-t_2)\hat{H})$ is the evolution operator.
The control variables $u$ now parametrize an initial state, and not
the Hamiltonian $\hat{H}$. But they also determine the evolution of
the system, $u \to z_u$, so that one may pose the problem of the
optimisation of a functional $\tilde{F}(z_u(T),u)$. Although it is an
apparently different formulation of the problem to the previous one,
it can however be transformed into it via the change of
variable:
\begin{equation}
y(t) = z(t) - z^0(u)\,,
\end{equation}
that transforms Eqs.~(\ref{eq:zsys1}-\ref{eq:zsys2}) into:
\begin{eqnarray}
\dot{y}(t) & = & -{\rm i}\hat{H}y(t) - {\rm i}\hat{H}z^0(u)\,,
\\
y(0) & = & 0\,,
\end{eqnarray}
i.e. they have the structure of
Eqs.~(\ref{eq:ysys1}-\ref{eq:ysys2}). One may then employ the usual
machinery of QOCT in order to arrive, for example, at an equation
for the gradient of the function $G(u) = \tilde{F}(z_u(T),u)$:
\begin{eqnarray}
\nonumber
\frac{\partial G}{\partial u_k} & = &
2{\rm Re}
\langle \frac{\partial \tilde{F}}{\partial z^*}(z_u(T),u)\vert\hat{U}(T,0)
\vert \frac{\partial z^0}{\partial u_k}\rangle
\\
\label{eq:grad}
& & +\left.\frac{\partial \tilde{F}}{\partial u_k}(z(T),u)\right|_{z(T)=z_u(T)}
\,.
\end{eqnarray}
Details of the derivation of this equation are given in the appendix.

The previous result is rather general. It may be simplified if we assume that (i)
the target functional takes the form:
\begin{equation}
\tilde{F}(z(T),u) = \langle z(T) \vert \hat{O} \vert z(T) \rangle\,,
\end{equation}
where $\hat{O}$ is some observable (e.g. projection onto some subspace),
and (ii) the variables $u$ are the (complex)
expansion coefficients of the initial state in terms of the orthonormal basis of
some subspace:
\begin{equation}
z^0(u) = \sum_i u_i \psi_i\,.
\end{equation}
The function to optimise is then:
\begin{equation}
G(u) = \langle z^0(u) \vert \hat{U}(0,T) \hat{O} \hat{U}(T,0) \vert z^0(u)\rangle\,,
\end{equation}
and due to both the linearity of $\hat{U}$ and the linearity of the previous
definition of $z^0(u)$ with respect to $u$, the function
$G$ may then be written as a quadratic form:
\begin{equation}
G(u) = \sum_{ij} u^*_i u_j \lambda_{ij} = u^\dagger {\bf \lambda} u\,,
\end{equation}
where
\begin{equation}
\label{eq:matrixlambda}
\lambda_{ij} = 
\langle \psi_i \vert \hat{U}(0,T) \hat{O} \hat{U}(T,0) \vert \psi_j\rangle\,.
\end{equation}
The search for an optimal $u$ cannot be unconstrained, since we must
assume that the initial state is normalised, i.e. $u^\dagger u =
1$. A quadratic form constrained in such a way has its critical points
at its eigenvectors:
\begin{equation}
\lambda u = \mu u\,.
\end{equation}
One may then choose the largest eigenvalue $\mu_0$ and its corresponding
eigenvector $u_0$, as the solution of the maximisation problem:
\begin{equation}
G_0 = G(u_0) = \max_{u^\dagger u=1} \langle z_u(T)\vert \hat{O} \vert z_u(T)\rangle\,.
\end{equation}
Numerically, the cost of the algorithm amounts to the cost of
propagating each basis element $\psi_j$ separately in order to compute
the matrix elements [Eq.~\ref{eq:matrixlambda}], plus the cost of the
matrix diagonalisation. This is the procedure that we will use in the
examples below.

\section{\label{sec:model} Model}

The Dicke Hamiltonian~\cite{Dicke1954,Garraway2011} models $N$ atoms
(in fact, generically speaking, TLSs of whatever origin) interacting
with a single mode of the radiation field (the extension to more than
one mode will discussed later on). It can be considered as a
generalisation of the Rabi model, for which $N=1$. It is given by:
\begin{equation}
\hat{H} = \omega \hat{a}^\dagger\hat{a} + \omega_0 \hat{J}_z 
+ g (\hat{a}^\dagger + \hat{a}) (\hat{J}_+ + \hat{J}_-)\,.
\end{equation}
It describes a single cavity mode (whose frequency is $\omega$, and whose
creation and annihilation operators are $\hat{a}^\dagger$ and
$\hat{a}$, respectively), coupled to a set of $N$ TLSs. The
$\hat{J}_z$ and $\hat{J}_\pm$ operators are the collective operators
describing those:
\begin{equation}
\hat{J}_z = \frac{1}{2}\sum_i^N \hat{\sigma}_z^{(i)}
\end{equation}
is the ``population operator'', that sums all the Pauli $\sigma_z$ operators
acting on each TLS $i$ (whose energy is $\omega_0$, although in the
following we will always assume the resonance condition, $\omega_0 = \omega$), and
\begin{equation}
\hat{J}_{\pm} = \frac{1}{2}\sum_i^N \hat{\sigma}_\pm^{(i)}
\end{equation}
are the collective ``ladder'' operators. The basis states for each TLS are
$\vert\downarrow\rangle$ and $\vert\uparrow\rangle$. The constant $g$ is the
atom-cavity coupling constant. Assuming that the system is truly
closed (all possible decay rates are negligible), the ratio
$g/\omega$ determines whether (i) the model operates in the regime of
validity of the rotating wave approximation (RWA), that permits
to ignore the counter-rotating terms ($\hat{a}\hat{J}_-$ and $\hat{a}^\dagger\hat{J}_+$), i.e.:
\begin{equation}
\label{eq:taviscummings}
\hat{H}\approx\hat{H}_{\rm TC} = \omega \hat{a}^\dagger\hat{a} + \omega_0 \hat{J}_z 
+ \frac{1}{2} g (\hat{a}^\dagger\hat{J}_- + \hat{a}\hat{J}_+)\,,
\end{equation}
or (ii) the model operates in the ultrastrong regime and the full
Hamiltonian has to be considered. In Eq.~(\ref{eq:taviscummings}), ``TC'' stands
for Tavis and Cummings~\cite{Tavis1968,Tavis1969}. For $N=1$, the TC model
reduces to the analytically solvable Jaynes-Cummings model~\cite{Jaynes1963,Shore1993}.

The coupling constant $g$ has another physical meaning: it is the Rabi
frequency for this latter Jaynes-Cummings model: a system departing
from the $\vert\downarrow\rangle\otimes\vert n\rangle$ state oscillates,
with frequency $n^{1/2} g$, between this and the
$\vert\uparrow\rangle\otimes\vert n-1\rangle$ state (where $\vert n\rangle$
and $\vert n-1\rangle$ are the photon Fock states with $n$ and $n-1$
photons, respectively).

Various physical systems have been found to be good candidates to
realise the Dicke model. For example, the experiments on
Rydberg-excited atoms passing through cavities~\cite{Raimond2001};
although many of these experiments have focused on the single atom and
on the \emph{micromaser}~\cite{Meschede1985} concept, collective
multi-atom effects have also been observed~\cite{Raimond1982}.  A
different frequency regime is that of optical cavities, which may also
be well described by the Dicke Hamiltonian in appropriate
circumstances~\cite{Agarwal1998}. Both the Rydberg atom and the optical cavity setup, however,
typically have a flux of atoms, and not a constant number as one tends
to assume theoretically. This inconvenience is absent in the ion
traps~\cite{Harkonen2009} experiments -- although the decay rates can be
high. Finally, we mention circuit QED, that places superconducting
qubits (instead of atoms) inside waveguides at microwave frequencies;
the Dicke model can also be used for these systems~\cite{Fink2009}.

In the next section, we will show how the preparation of some initial
states may lead to the creation of other target states -- that could
in turn be used as initial states for subsequent processes. For that purpose,
we have focused on two sets of target states:
\begin{itemize}

\item Fock states, also known as number states, are perhaps the most
  ``quantum'' states of light (and of harmonic oscillators in general), as
  they are the states with a definite number of light quanta,
  i.e. photons. These states have potential applications in quantum
  metrology, cryptography and computing. Unfortunately, the creation
  of Fock states is challenging, and numerous protocols have been
  proposed, e.g. Refs.~\cite{FrancaSantos2001,Bertet2002,Brown2003,Zhang2012}, or
  Ref.~\cite{Hofheinz2008} for Fock states in a superconducting
  quantum circuit.

\item Given a number $N$ of TLSs, the Dicke states~\cite{Dicke1954} are defined as:
\begin{equation}
\vert D(N, k)\rangle = {N\choose k}^{-\frac{1}{2}} {\rm sym}
\left[
\vert \downarrow \rangle^{\otimes N-k}
\otimes
\vert \uparrow \rangle^{\otimes k}
\right]\,,
\end{equation}
where ${\rm sym}$ means symmetrisation. A Dicke state is therefore the
equal superposition of all basis states of $N$ qubits having exactly
$k$ excitations ($\vert\uparrow\rangle$ states). For the case $k=1$ (only one excitation), the Dicke
states are also called $W$-states~\cite{Dur2000}: $W(N) = D(N,1)$. For example,
assuming a TLS set with three systems ($N=3$):
\begin{equation}
\vert W(3) \rangle = 3^{-1/2} \left[
\vert \uparrow \rangle \otimes \vert \downarrow \rangle \otimes \vert \downarrow \rangle +
\vert \downarrow \rangle \otimes \vert \uparrow \rangle \otimes \vert \downarrow \rangle +
\vert \downarrow \rangle \otimes \vert \downarrow \rangle \otimes \vert \uparrow \rangle
\right]\,.
\end{equation}
Dicke states play an important role in quantum optics and quantum
information theories, due to their entanglement and
nonlocality~\cite{Barnea2015}. They have been 
prepared experimentally in various ways -- see for example
Ref.~\cite{Wu2017} and references therein.

\end{itemize}

\section{\label{sec:results} Results}

\subsection{Creation of one-photon Fock states}

We start with a simple example: the goal is to create the one photon
Fock number state in an empty cavity, by tailoring the initial TLS set state
that enters at time zero. Therefore, the initial parametrization of the 
initial state is given by:
\begin{equation}
\vert z^0(u) \rangle = \sum_i u_i \vert \phi_i \rangle \otimes \vert 0 \rangle
\end{equation}
where $i$ runs over all the possible TLS set basis states $\vert \phi_i \rangle$,
and $\vert 0\rangle$ is the photon vacuum. The operator $\hat{O}$ that
determines the target is in this case given by:
\begin{equation}
\hat{O} = \hat{I}_{\rm M} \otimes \vert 1 \rangle \langle 1 \vert\,.
\end{equation}
where $\hat{I}_{\rm M}$ is the identity operator in the ``matter''
space (the set of TLSs).

We have first set the coupling constant $g$ to a small value,
well within the RWA validity range (weak coupling regime): $g/\omega = 10^{-6}$, consistent for example
with the experimental setups described in Ref.~\cite{Raimond2001}. For
$N=1$ TLS, we would therefore have the Jaynes-Cummings Hamiltonian,
whose evolution is analytically known, and the optimisation problem
could be solved without any calculation: if the TLS enters the system
in its excited state $\vert\uparrow\rangle$, the full system
oscillates from $\vert\uparrow\rangle\otimes\vert 0\rangle$ to
$\vert\downarrow\rangle\otimes\vert 1\rangle$ and back with frequency
$g$ and period $\tau = \frac{2\pi}{g}$. If the total time is then set to half
that period, the target state achieves full
population for the $\vert\uparrow\rangle\otimes\vert 0\rangle$ initial
state.

\begin{figure}
\centerline{\includegraphics{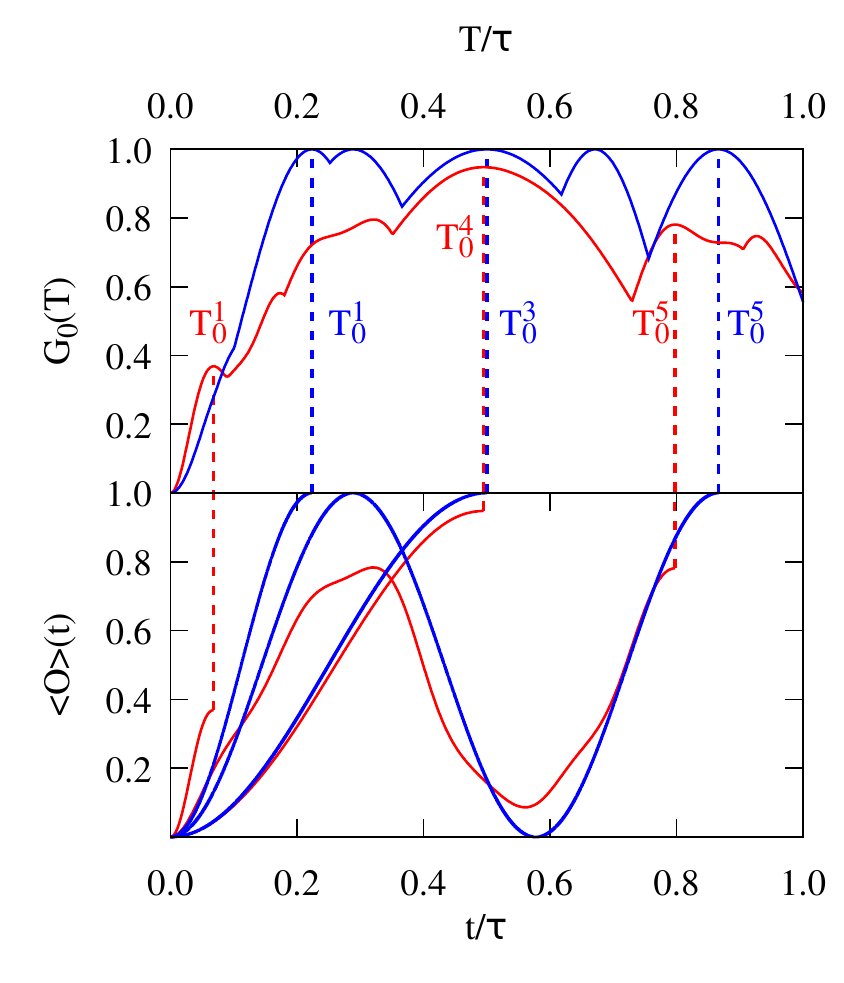}}
\caption{
\label{fig1}
{\it Top:} Optimal occupation of the one-photon Fock number state, as a function
of the total interaction time $T$ for a weak coupling case ($g \ll \omega$, blue), and
an ultrastrong coupling case ($g \approx 0.5 \omega$, red). {\it Bottom:} Evolution in time of the
one-photon Fock number state occupation, when the initial state is optimal
for various interaction times $T_0^j$ corresponding to local maxima of the $G_0(T)$ curve.
($j$ indexes the maxima in increasing time order).
}
\end{figure}

This full population can only be achieved by fixing $T$ to that
transition time, or odd multiples of it. At any other time, the
optimal population would be lower, as it strongly depends on $T$. This
fact also holds for larger $N$ values: for example, in Fig.~\ref{fig1}
(top) we display the optimal population of the one-photon Fock state
obtained when performing optimisations at varying values of $T$, for
the $N=5$ case. The blue data in Fig.~\ref{fig1} corresponds to the
results in the weak coupling regime. One may see how the full
population can only be achieved at certain values of $T$. This fact
suggest adding $T$ to the set of control variables, as it is a
parameter that can often be controlled in experimental realisations of
these models. This could be formally done generalising the previous
equations from the beginning, but in our case we have preferred to do
two consecutive searches: first, optimisations at fixed $T$ values as
described above in order to build a $G_0(T)$ curve, and then one
search for the maxima of this curve (named $T_0^j$ in the plot, where
the index $j$ orders the local maxima in time).

For example, one could be interested in finding the fastest way to
create the one-photon state. This would be achieved by using the
optimal initial state that produces the first local maximum of the
curve shown in Fig.~\ref{fig1}, i.e. $T_0^1$. The evolution of
$\langle \hat{O}\rangle(t)$ for the various optimal initial states are
depicted in Fig.~\ref{fig1} (bottom, blue lines for the results
obtained within the weak coupling regime).

By repeating this process with varying values of number of TLSs $N$,
we found that: (i) in all cases, the one-photon Fock state could be
produced exactly; (ii) the optimal initial state corresponding to the
fastest transition is given by the $W(N)$ state; and (iii) the optimal
interaction time follows a simple rule $T_0^1(N) =
\frac{1}{2}N^{-1/2}\tau$. The time required for the one-photon
state to be created is smaller as the number of TLSs grows.

We next explored in the same way the ultrastrong regime, by doing the
same optimisation runs with values of $g \approx 0.5\omega$.  The
results are displayed in red in Fig.~\ref{fig1}.  The findings above
do not hold: (i) in the ultrastrong regime, the perfect creation of
the one photon state is not achieved (the best local maxima is
$\approx$ 0.94), (ii) the optimal initial state corresponding to the
fastest optimal transition is no longer a $W$-state, and (iii) the
optimal transition time no longer follows the simple rule given above.
These results can be explained: the $W$-state is a linear combination
of ``single-excitations'' of the TLS set, that transforms in the weak
coupling regime into a one-photon state, a fact that is related to a
conservation law (the number of matter excitations, plus the number of
photons), valid within this regime. As the coupling becomes stronger,
the optimal initial state has components of TLS states with two and
more excitations.

\subsection{Creation of $k$-photon Fock states}

\begin{figure}
\centerline{\includegraphics{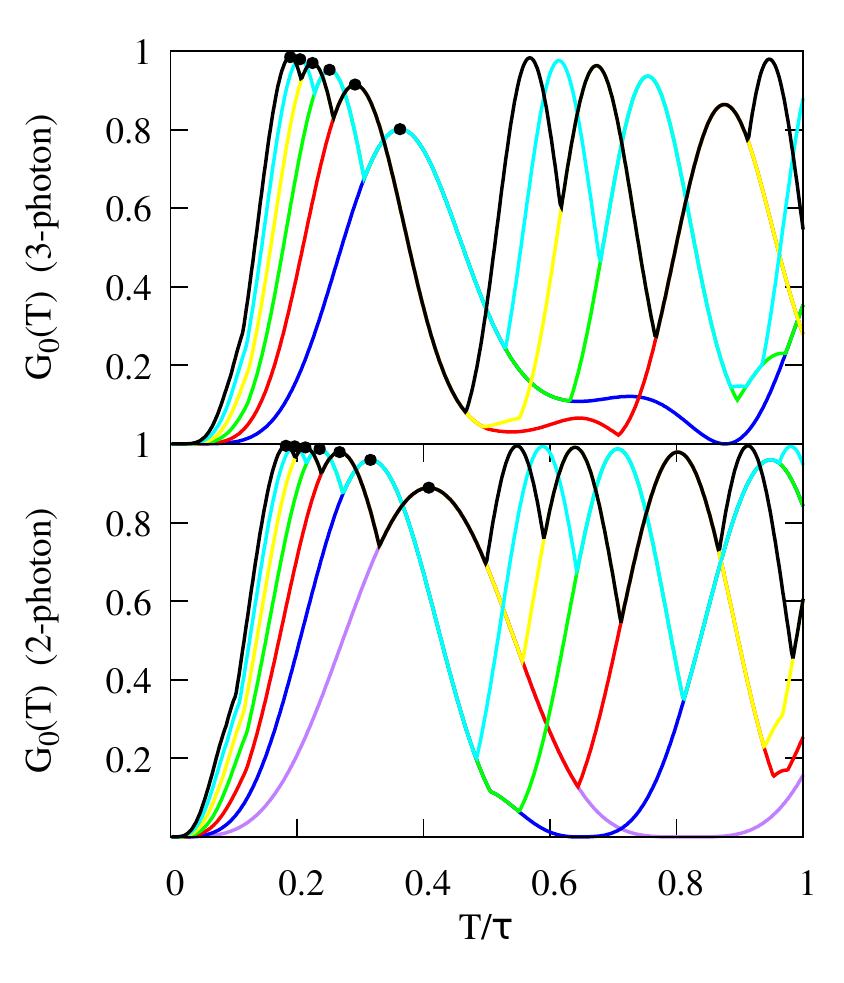}}
\caption{
\label{fig2}
Optimal occupation of the two-photon (top) and three-photon (bottom) Fock number states,
as a function of the total interaction time $T$, for various values of the
number of TLSs: violet (2), blue (3), red(4), green (5), yellow (6), cyan (7) and
black (8). Note that often the curves overlap. The black dots mark the first local
maxima.
}
\end{figure}

The same procedure can be used to create Fock states with a larger
number of photons. For example, in Fig.~\ref{fig2} we show the results
obtained when using the two-photon and three-photon states as
targets, i.e.:
\begin{equation}
\hat{O} = \hat{I}_{\rm M} \otimes \vert k \rangle \langle k \vert\,,
\end{equation}
for $k=2,3$. The plot displays the optimal value attained with varying
values of the total interaction time, for several numbers of
TLSs. These calculations were performed in the already introduced weak
coupling regime.

First, note that the target state is never created with perfect
fidelity, in contrast to the one-photon case. However, by increasing
$N$, the fidelity grows and reaches values that get arbitrarily
close to one. This can be seen by looking for example at the first
local maxima of the curves, displayed in the plots with black dots.

One interesting point of those first local maxima is that they somehow
generalise the result of the previous section: the optimal initial
states of those local maxima are the Dicke states $D(N,k)$, with $k=2$
and 3 excitations for the 2-photon and 3-photon targets,
respectively. We observed this fact to hold for larger photonic
numbers: the optimal state for the fastest creation of the $k$-photon
Fock space is the $D(N,k)$ state, but the state is not created exactly
-- only in the limit of very large number of TLSs the fidelity can be
made arbitrarily close to one.

Finally, the optimal interaction times $T_0^1(N)$ are also reduced
with increasing $N$, following $N^{-1/2}\tau$ trends: $\frac{1}{\sqrt{2}}
(N-1)^{-1/2}\tau$ for the 2-photon target, and $\propto
(9N-10)^{-1/2}\tau$ for the 3-photon target.

\subsection{Creation of Dicke states}

We now turn to the more familiar problem in the field of QOCT: the
attempt to shape an electro-magnetic field such that it induces a
given behaviour on a piece of matter, e.g. the preparation of some
matter state. Given the special role played by Dicke states when attempting
to prepare Fock states, we will set them now as targets for the
optimisations. This is achieved by setting the target operator as:
\begin{equation}
\hat{O} = \vert D(N,k)\rangle  \langle D(N,k)\vert \otimes \hat{I}_{\rm F}\,,
\end{equation}
where $\hat{I}_{\rm F}$ is the identity operator in photon space, $N$
is the number of TLSs present in the cavity, and $k$ is the number of
excited TLSs in the definition of the state. The TLSs are assumed to 
be in their ground state at time zero, and therefore the parametrization
of the initial state is in this case:
\begin{equation}
\vert z^0(u) \rangle = \sum_i^M u_i \left[ \vert \downarrow \rangle^{\otimes N} \right]
\otimes \vert i \rangle\,,
\end{equation}
where $M$ is the cut-off number of photons included in the calculation
(a number that we carefully checked to be big enough not to affect the
results).

\begin{figure}
\centerline{\includegraphics{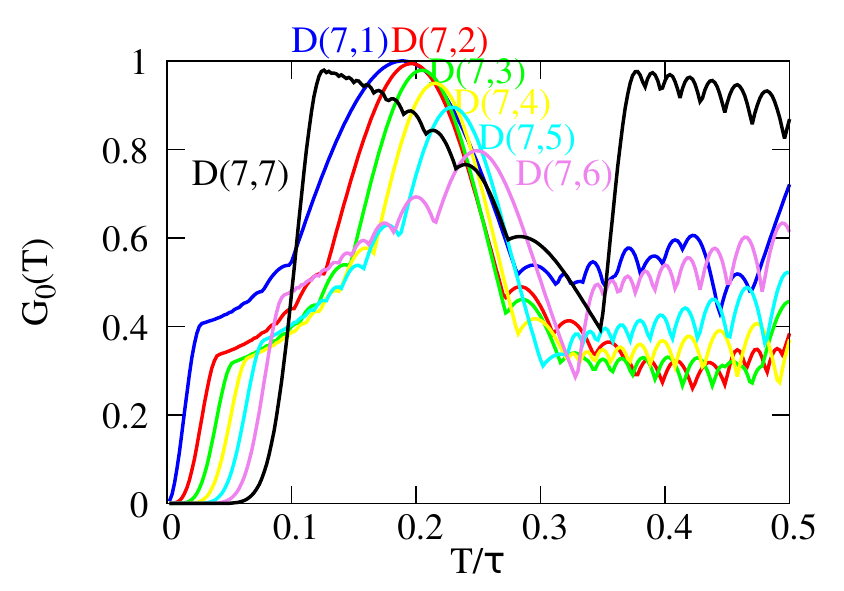}}
\caption{
\label{fig3}
Optimal occupation of the marked Dicke states, as a function of the total interaction
time $T$.
}
\end{figure}

As an example, we display in Fig.~\ref{fig3} the optimisations
achieved for the $D(7,k)\;\;(k=1,\dots,7)$ series. The best possible
Dicke states will be created at the interaction times that determine
the maxima of the curves. Unsurprisingly, the optimal initial states
that correspond to the first maxima are, precisely, the Fock number states
with $k$ photons. It becomes clear that Fock number states and Dicke states
play a ``conjugated'' role: Fock states are the optimal initial states
for the creation of Dicke states, and vice versa.

By looking at the first maxima of the curves one may also note that
(i) the fidelity in the production of the Dicke states deteriorates
with increasing $k$ -- in fact, only the $W$-state ($k=1$) can be
created exactly; and (ii) the time needed for the (at least
approximate) creation of the Dicke states also becomes longer with
increasing $k$. The exception is the $D(7,7)$ state, whose maximum
does not correspond to an initial Fock number state but to a linear combination of those, and whose optimal
interaction time does not follow the trend of the other cases. The
$D(7,7)$ state is indeed peculiar, as it is not an entangled state,
but the product of all the TLSs excited states.

\subsection{Multi-mode photon states}

The previous examples have assumed a single cavity mode; an obvious
generalisation of the Dicke Hamiltonian is the consideration of a
number $L$ of radiation modes:
\begin{eqnarray}
\nonumber
\hat{H} & = & \sum_n^L \omega_n \hat{a}_n^\dagger\hat{a}_n + \frac{1}{2}\sum^N_k \omega^{(k)}_0 \hat{\sigma}_z 
\\
& & + \sum_n^L \sum_k^N g_n^{(k)} (\hat{a}_n^\dagger + \hat{a}_n) (\hat{\sigma}^{(j)}_+ + \hat{\sigma}^{(+)}_-)\,.
\end{eqnarray}
The extra freedom permits to attempt the creation of more complex photon states, such as,
if we work with two modes:
\begin{equation}
\vert R_1 \rangle = \frac{1}{\sqrt{2}}\left( \vert 10\rangle + \vert 01\rangle\right)\,,
\end{equation}
as an example of one-photon state, and, if we work with three modes and two-photon states:
\begin{equation}
\vert R_2 \rangle = \frac{1}{\sqrt{2}}\left( \vert 110\rangle + \vert 101\rangle \right)\,.
\end{equation}
The corresponding target operators are $\hat{O}=\hat{I}_M\otimes \vert
R_\mu\rangle\langle R_\mu\vert$ ($\mu=1,2$). The first case involves
the presence of (at least) two modes, whereas the second case requires
three. The photons occupying these modes are thus entangled -- the
$R_1$ is the prototypical (and controversial) example of single-particle entangled
state~\cite{vanEnk2005,Drezet2006,vanEnk2006}.

In order to successfully couple to those modes, we need to have a
number of TLSs resonant to the corresponding frequencies, i.e. some of
the $w_0^{(k)}$ must be equal to each $\omega_n$. We therefore set a
number $N_n$ of TLSs resonant to each mode ($\sum_n N_n = N$) and then
perform the optimisation using the same procedure as in the previous
cases.

\begin{figure}
\centerline{\includegraphics{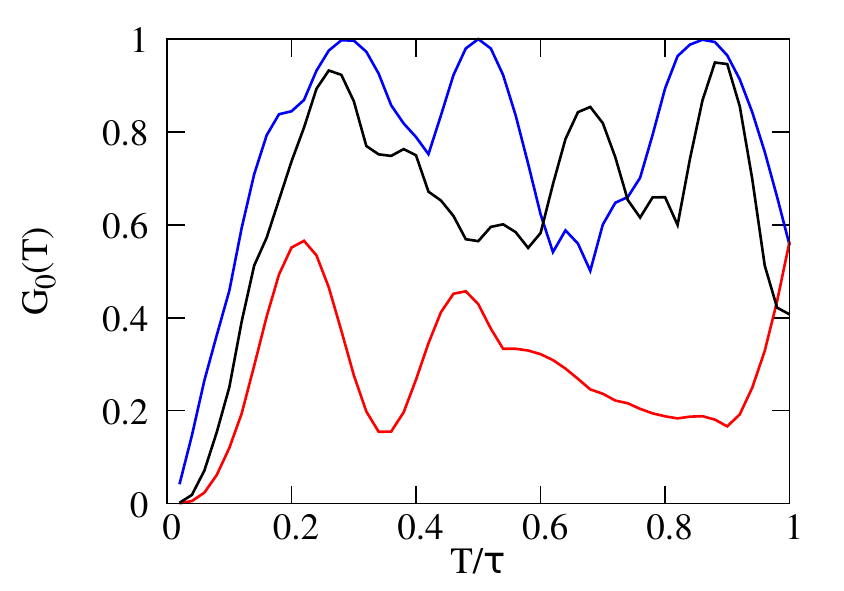}}
\caption{
\label{fig4}
Optimal occupation of the $R_1$ as a function of the interaction time
$T$ in the presence of three TLS per mode (blue). Optimal occupation of the $R_2$ state
in the presence of two (red) and three (black) TLSs per mode.}
\end{figure}

The results are shown in Fig.~\ref{fig4}. We plot the optimal
occupation of the states as a function of the total interaction time
$T$, for both targets. For $R_1$, the plot shown (blue curve) is the
one corresponding to three TLSs per mode ($N_1=N_2=3$), and it can be
seen how there are several optimal times for which the achieved occupation
is one.  For $R_2$ we display two sample calculations, one
with two TLSs per mode (red) and one with three (black). It can be
seen how the optimal occupations achieved with the latter are higher,
and therefore we have the expected trend: the more TLSs, the more
freedom in the variational search, and therefore the better the results.

\section{\label{sec:conclusions} Conclusions}

The machinery of QOCT can be easily extended to include the problem of
the optimisation of the initial state of a quantum process such that
its evolution produces a pre-defined target outcome -- such as the
creation of a given state, the maximisation of the projection onto a
Hilbert subspace, etc. This idea may be used to generalise the typical
QOCT problem -- the shaping of an external electromagnetic field in
order to control the evolution of a quantum system -- to include the
case in which the field is no longer external but a part of the system
itself. In this case, the shaping must be understood as the
preparation of the initial state of the field. Immediately, this idea
suggests the complementary concept of shaping the initial state of the
matter subsystem in order to control the subsequent evolution of the
electromagnetic quantum state.

As in standard QOCT, the method starts with the definition of a target
functional. The crucial equations are those of the gradient of this
functional with respect to the parameters that define, in this case,
the shape of the initial state. The calculation of this gradient
essentially requires multiple propagations of the system in
time. Therefore, the computational complexity depends on the cost of
these propagations. If (1) the target functional is the expectation
value of the final state on some operator; (2) the control variables
are the expansion coefficients of the initial state in some subspace
of allowed initial states; and (3) the quantum evolution equation is
the linear Schr{\"o}dinger equation, then the problem is reduced to
the maximisation of a quadratic form, which amounts to
the diagonalisation of a matrix whose coefficients are computed by
propagation of the basis states.

We have tested these ideas using the Dicke model -- a set of TLSs
coupled to one radiation mode (or more, in a generalised version) in a
cavity.  We have examined the creation of Dicke states through
the initial preparation of the electromagnetic state, and the creation
of Fock radiation states through the initial preparation of the states
of the TLSs. Dicke and Fock number states play a conjugated role, as Dicke
states turn out to be the optimal initial states for the preparation
of Fock number states, and vice versa. The creation of the target states is
unfortunately not always exact, and depends on the size of the initial
state search subspace -- for example, Fock number states are better created
if more TLSs are included in the model.

Finally, we outlook a few possible extensions and modifications of
this work: (1) First, we note that this reformulation of QOCT is not
incompatible with the usual one, as one could have the freedom of
shaping both the initial state and an external field -- the equations
shown above could be trivially extended to include this option. This
could be used to guide experimental work in which one has the
possibility of both partially preparing the initial state and acting
externally on the subsequent evolution. (2) Second, one could take the
classical or mean field limit for the electromagnetic field, and
consider an OCT formulation for the coupled Maxwell-Schr\"odinger
dynamics. This OCT should therefore target a mixed quantum-classical
system, a possibility that has already been realised for models with
quantum electrons and classical nuclei~\cite{Castro2014}. (3) In this
work, the matter subsystem has been modelled with simplified two-level
Hamiltonian, but an ab-initio coupled treatment of many-particle systems
and photon fields is possible -- using, for example, the density
functional reformulation of the quantum electrodynamics
equations~\cite{Flick2015}. Work along these three lines is in
progress.

\begin{acknowledgments}
AC acknowledges support from the MINECO FIS2017-82426-P grant. We
acknowledge financial support from the European Research Council
(ERC-2015-AdG-694097) and Grupos Consolidados (IT578-13). The Flatiron
Institute is a division of the Simons Foundation.
\end{acknowledgments}

\appendix

\section{\label{sec:appendix} Derivation of the control equations}

The control problem formulated through
Eqs.~(\ref{eq:ysys1}-\ref{eq:ysys2}) and Eq.~(\ref{eq:g1}) may be approached through 
Pontryagin's principle~\cite{Boltyanskii1956,Pontryagin1962}, that
establishes a set of necessary conditions for the solution. 
In essence, the solution must be found at one of the zeros of the
gradient of $G$, that is given by:
\begin{eqnarray}
\nonumber
\label{eq:grad1}
\frac{\partial G}{\partial u_k} & = &
\left.\frac{\partial F}{\partial u_k}(y(T),u)\right|_{y(T)=y_u(T)} + 
\\
& & 2{\rm Re} \int_0^T\!\!{\rm d}t\;\langle\lambda_u(t)\vert
\left[ -{\rm i}\frac{\partial \hat{H}(u,t)}{\partial u_k}\vert y_u(t)\rangle 
+ \vert \frac{\partial b}{\partial u_k}(u,t)\rangle
\right]
\end{eqnarray}
In this equation, the \emph{costate} is defined as the solution to:
\begin{eqnarray}
  \dot{\lambda}_u(t) & = & -{\rm i}\hat{H}^\dagger(u,t)\lambda_u(t)\,,
  \\
  \lambda(T) & = & \frac{\partial F}{\partial y^*}(y_u(T),u)\,.
\end{eqnarray}

Now we re-consider the reformulation of the problem for the system described through 
Eqs.~(\ref{eq:zsys1}-\ref{eq:zsys2}), i.e.:
\begin{eqnarray}
\dot{z}(t) & = & -{\rm i}\hat{H}z(t)\,,
\\
z(0) & = & z^0(u)\,,
\end{eqnarray}
for a target function $\tilde{F}(z(T),u)$. Upon
the change of variable $y(t) = z(t) - z^0(u)$, we get:
\begin{eqnarray}
\dot{y}(t) & = & -{\rm i}\hat{H}y(t) - {\rm i}\hat{H}z^0(u)\,,
\\
y(0) & = & 0\,.
\end{eqnarray}
We may therefore use Eq.~(\ref{eq:grad1}), which results in:
\begin{eqnarray}
\nonumber
\frac{\partial G}{\partial u_k} & = &
2{\rm Re} \langle \frac{\partial \tilde{F}}{\partial z^*}(z_u(T),u)\vert 
\frac{\partial z^0}{\partial u_k}\rangle + 
\left.\frac{\partial \tilde{F}}{\partial u_k}(z(T),u)\right|_{z(T)=z_u(T)}
\\\label{eq:app3int1}
& & + 2{\rm Re}\left[ -{\rm i}\int_0^T\!\!{\rm d}t\;\langle\lambda_u(t)\vert
H \vert \frac{\partial z^0}{\partial u_k}\rangle\right]
\,.
\end{eqnarray}
Now the \emph{costate} $\lambda_u$ is given by::
\begin{equation}
\lambda_u(t) = \hat{U}(t,T)\frac{\partial \tilde{F}}{\partial z^*}(z_u(T),u)\,.
\end{equation}
Therefore:
\begin{equation}
\int_0^T\!\!{\rm d}t\;\langle\lambda_u(t)\vert H = 
\int_0^T\!\!{\rm d}t\;\langle \frac{\partial \tilde{F}}{\partial z^*}(z_u(T),u) \vert U(T, t)H
\end{equation}
If we use $U(T, t)H = -i \frac{\rm d}{{\rm d}t}U(T,t)$,
we finally arrive at:
\begin{equation}
\label{eq:grad2}
\frac{\partial G}{\partial u_k} =
\left.\frac{\partial \tilde{F}}{\partial u_k}(z(T),u)\right|_{z(T)=z_u(T)}
+2{\rm Re}
\langle \lambda_u(0)
\vert \frac{\partial z^0}{\partial u_k}\rangle\,.
\end{equation}
This equation may also be rewritten without any reference to the costate,
as in Eq.~(\ref{eq:grad}).

\bibliography{qoctqed}

\end{document}